\documentclass[aps, prb, showpacs, twocolumn, superscriptaddress, floatfix]{revtex4-1}
\usepackage{graphicx}
\usepackage{subfigure}
\usepackage[breaklinks=true]{hyperref}
\usepackage{breakcites}
\begin{document}

%\preprint{}

\title{Testing Predictions from Density Functional Theory at Finite Temperatures: $\beta_2$-Like Ground States in Co-Pt}
% \title{Shortcomings of Density Functional Theory in Transition Metal Intermetallics: $\beta_2$-Like Ground States Predicted in Co-Pt}

\author{Elizabeth Decolvenaere}
%\email{elizabeth@umail.ucsb.edu}
\affiliation{Department of Chemical Engineering, University of California Santa Barbara, Santa Barbara, California 93106, USA}

\author{Michael J. Gordon}
%\email{mjgordon@engineering.ucsb.edu}
\affiliation{Department of Chemical Engineering, University of California Santa Barbara, Santa Barbara, California 93106, USA}

\author{Anton Van der Ven}
\email{avdv@engineering.ucsb.edu}
\affiliation{Materials Department, University of California Santa Barbara, Santa Barbara, California 93106, USA}


\date{\today}

\begin{abstract}
   We perform a critical assessment of the accuracy of DFT-based methods in predicting stable phases within the Co-Pt binary alloy. Statistical mechanical analysis applied to zero kelvin DFT predictions yields finite-temperature results that can be directly compared with experimental measurements. The predicted temperature-composition phase diagram is qualitatively incompatible with experimental observations, indicating that the predicted stability of long-period superstructures as ground states in the Co-Pt binary is incorrect. We also show that recently suggested methods to better align DFT and experiment via the hybrid functional HSE06 are unable to resolve the discrepancies in this system. Our results indicate a need for better verification of DFT based phase stability predictions, and highlight fundamental flaws in the ability of DFT to treat late 3\textit{d}-5\textit{d} binary alloys.
\end{abstract}

%\keywords{}

\maketitle


\section{Introduction}

Density functional theory (DFT) has grown to become the most popular electronic structure calculation method to date\cite{Becke2014}. Modern computational resources have made DFT viable as a high-throughput materials design technique\cite{Jain2011, Jain2013, Curtarolo2013}, whereby the existence, stability, and properties of periodic crystalline phases are predicted entirely from first principles. These approaches are especially attractive for predicting the properties of systems that are otherwise too expensive or difficult to study experimentally, such as alloys containing Ru\cite{Jahnatek2011}, Tc\cite{Levy2012}, and Pt\cite{Hart2014}, among others\cite{Curtarolo2005,Ghosh2008,Levy2010, Maisel2012}.

While remarkably successful in predicting phase stability in a wide variety of chemically disparate systems\cite{Curtarolo2005}, the occasional failures of DFT\cite{Feibelman2001, Zhou2004, Zhang2014} highlight the importance of experiments to validate such predictions. However, there exists a fundamental challenge in comparing DFT and experiments: electronic structure calculations predict zero kelvin properties, while experiments are performed at finite temperatures. The most reliable measurements of thermodynamic properties are performed at elevated temperatures, where equilibrium is more readily attained, but also where the entropic contributions to such properties are the greatest. Accuracy in comparing \textit{ab-initio} and experimental results is vital, because any mismatch may indicate failure in the approximations used in DFT to accurately reproduce the necessary physics.

Of particular concern are disagreements between the set of observed phases and DFT-predicted zero kelvin ground states. These errors, while more sensitive to unknown kinetic barriers or unaccounted for entropic contributions, can also indicate fundamental flaws in the \textit{ab-initio} method. Binaries that pair late 3\textit{d} with late 5\textit{d} transition metals, such as Cu-Au\cite{Ozolins1997, Sanati2003}, Co-Pt\cite{Hart2014, Chepulskii2011}, Ni-Pt\cite{Sanati2003, Shang2011, Hart2014}, Fe-Pt\cite{Barabash2009, Hart2014}, and Fe-Pd\cite{Chepulskii2012}, represent one class of materials where DFT predicts a rich variety of zero kelvin ground states for which no experimental evidence exists. A wealth of long-period superstructures have been predicted to be stable in the intermediate continuum of compositions between $x_{Pt}$ 0.5 and 0.75 in the Ni$_{1-x}$Pt$_x$, Fe$_{1-x}$Pt$_x$, Co$_{1-x}$Pt$_x$ and Cu$_{1-x}$Au$_x$ alloys, instead of the two phase mixtures of L1$_0$ and L1$_2$ observed experimentally. In the case of Cu-Au, Co-Pt, Ni-Pt and Fe-Pd, the L1$_2$ AB$_3$ structure is altogether excluded from the set of ground states, and in the case of Co-Pt, the L1$_0$ formation enthalpy is less stable than the (experimental) solid solution enthalpy\cite{Kim2011}. In all cases, the formation enthalpies for the ordered phases have been predicted to be dozens of meV higher than experimental results\cite{Oriani1954, Oriani1962, Walker1970, Kessel2001}. These results derive from zero kelvin predictions, but little or no thermodynamic analysis of the finite-temperature impacts has been performed, and bulk phase diagrams deriving solely from electronic structure calculations have never previously been constructed for these materials. Co-Pt, Fe-Pt, and Fe-Pd are all candidates for use in ultrahigh density magnetic storage\cite{Coffey1995,Sato2000,Suess2005,Ferrando2008,Qin2009,Takenoiri2011}; resolving uncertainties about low-temperature predictions of phase (in)stability is thus critical.

Here, we explore the finite temperature implications of the zero kelvin ground states predicted by DFT for the Co$_{1-x}$Pt$_x$  alloy. We have developed a first-principles cluster-expansion Hamiltonian and used it in semi-grand canonical Monte Carlo simulations to construct a temperature versus composition phase diagram. The predicted phase stability is in qualitative disagreement with available experimental observations. Our calculations also indicate that long-range ordered phases persist as ground states when including corrections for spin-orbit coupling, antiferromagnetic and mixed magnetic ordering, and noncollinear magnetism. Using a hybrid functional, we have attempted to recover experimental results; however, this approach is shown to introduce new errors in the enthalpy and magnetic moments of the structures. These errors can be traced back to known failings of Hartree-Fock exchange when applied to transition metals, which we illustrate by analyzing the density of states (DOS) for L1$_0$ CoPt.

\section{Methods}
\subsection{Electronic Structure Calculations}

All electronic structure calculations were performed using the Vienna \textit{Ab Initio} Simulation Package (VASP)\cite{Kresse1994,Kresse1994,Kresse1996a,Kresse1996} with projector-augmented wave (PAW) potentials\cite{Kresse1999,Blochl1994a} using the PBE functional\cite{Perdew1996} for exchange and correlation (XC) energies\cite{Perdew1996}. In all cases, the energy cutoff for the plane wave basis set was 460 eV and a $\Gamma$-centered Monkhorst-Pack\cite{Monkhorst1976} \textit{k}-point mesh, converged to energy changes of less than 1 meV/atom, was used. All simulation cells were allowed to relax their volume, atomic positions, and collinear magnetic spins. Spin polarization was employed and the correlation interpolation formula of Vosko \textit{et al.}\cite{Vosko1980} was used to enhance magnetic energies.

\subsection{Cluster Expansion and Monte Carlo Simulations}
Configurational degrees of freedom in an alloy can be described with a cluster expansion Hamiltonian\cite{Sanchez1984, DeFontaine1994} combined with Monte Carlo simulations\cite{VanderVen2008,VanderVen2010, Puchala2013}. The fully relaxed energy [$E(\bar{\sigma})$] of the arrangement [$\bar{\sigma}$] of two components on a parent crystal structure can be expressed via the following Hamiltonian:\cite{Sanchez1984}
\begin{equation}
    E(\bar{\sigma}) = V_0 + \sum_{\alpha} V_\alpha \sum_{\delta \in \Omega_\alpha} \phi_\delta(\bar{\sigma}),
\end{equation}
where $\alpha$ indexes a type of cluster (e.g., a nearest-neighbor pair, a next-nearest-neighbor pair, a nearest-neighbor tetrahedron, etc.), $\Omega_\alpha$ is the \textit{set} of clusters symmetrically-equivalent to $\alpha$, and $\delta$ indexes different clusters in $\Omega_\alpha$. $\phi_\delta(\bar{\sigma})$ is the cluster function for the cluster indexed by $\delta$, and the $V_\alpha$ are effective cluster interactions (ECIs) obtained from projecting (e.g., via regression) \textit{ab initio} energies and configurations onto the basis set of clusters (with $V_0$ as the ``empty cluster'', i.e., a constant term). We chose values for $\sigma_i$ of -1 for Co and +1 for Pt, and $\phi_i(\bar{\sigma}) = \prod_{i \in \delta} \sigma_i$ (i.e., the product of the $\sigma_i$ of the sites in the cluster).

% \subsection{Monte Carlo Simulations}

Semi-grand canonical Monte Carlo simulations were performed in a $24 \times 24 \times 24$ periodic supercell with 1000 equilibration passes followed by 2000 thermodynamic-averaging passes (where one pass is $N_{sites}$ attempted ``spin flips''). Approximate phase boundaries were identified by discontinuities in the composition-temperature lines at constant chemical potential (using temperature increments of $\Delta T = 2$ K), or by plateaus in composition-chemical potential lines at constant temperature (using chemical potential increments of $\Delta \mu = 0.01$ eV).

\subsection{Phonons}
Force constants were calculated using the frozen phonon approach\cite{born1954, Parlinski1997, Wei1992}, perturbing large supercells (108 atoms) with small, isolated, symmetrically-distinct atomic displacements (0.01, 0.02, and 0.04 \AA). Following electronic structure calculations, the resulting force constants were used to construct the dynamical matrix\cite{Gonze1997}. Vibrational free energies\cite{vandeWalle2002} were then calculated using the quasi-harmonic approximation, repeating the previous procedure at a variety of slightly smaller and larger ($-2\%$ to $+10\%$) supercell volumes and using a second order polynomial to fit the dependence of the free energy on volume to determine the change in formation energy with temperature\cite{Ivy}.

\section{Results and Discussion}
\subsection{Zero kelvin and Finite Temperature Results for PBE}

\begin{figure}[h]
\subfigure{\label{hull} \includegraphics[width=0.45 \textwidth]{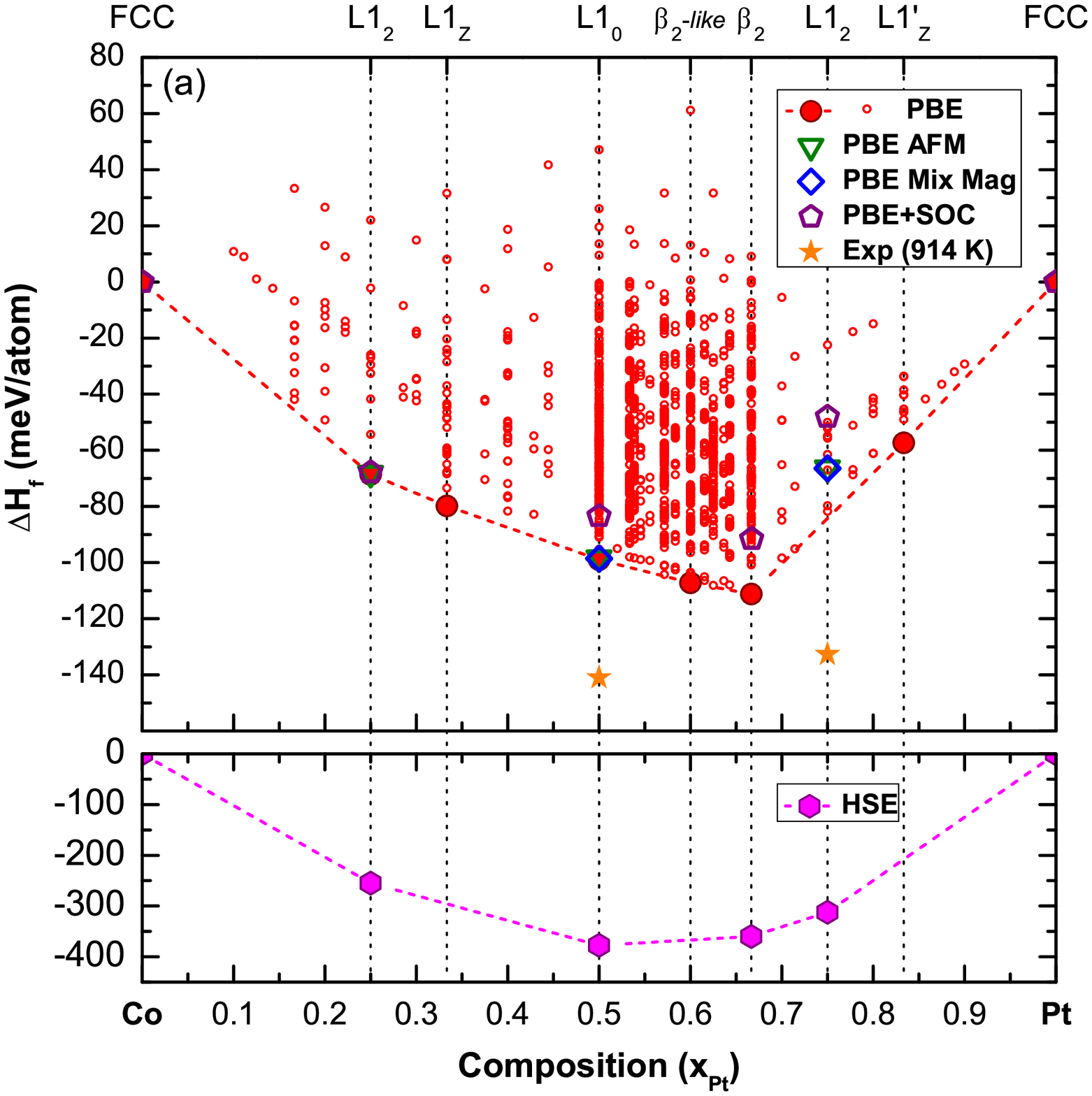}}
\subfigure{\label{struct} \includegraphics[width=0.45 \textwidth]{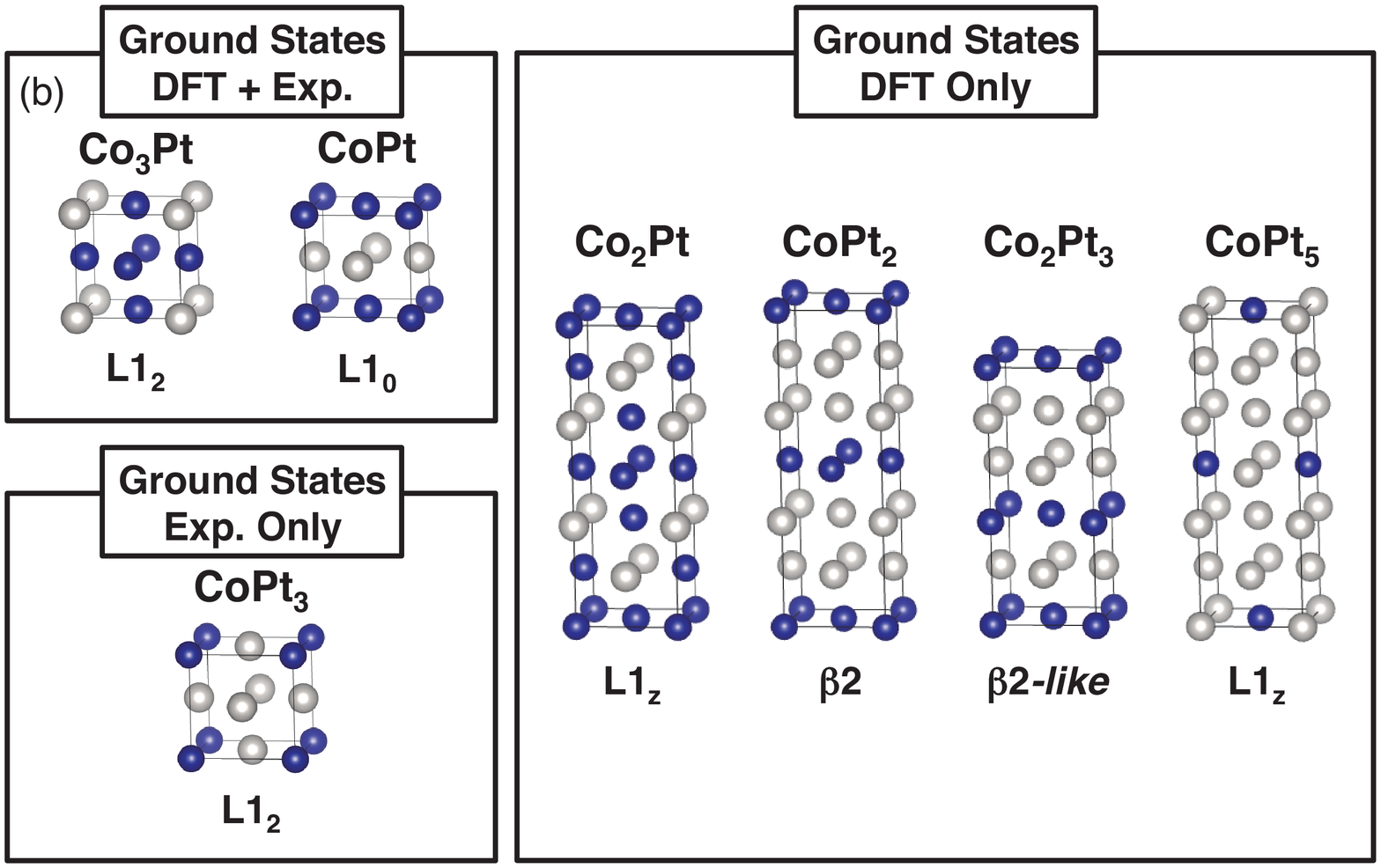}}
\caption{(color online) (a) Formation energies in the Co-Pt binary as calculated using PBE and HSE06. For ground state configurations, antiferromagnetic and mixed/ferrimagnetic orderings and SOC corrections were also considered. Experimental values at 914K are from Ref.~\onlinecite{Oriani1962}. (b) Crystal structures of the stable ground states predicted in Co-Pt using PBE (drawn with VESTA\cite{VESTA}). Compositions for ground states, including long-period superstructures, are indicated by dotted lines.\label{fig1}}
\end{figure}

To fully characterize the \textit{ab-initio} properties of the Co-Pt binary, we calculated the DFT energies of 1469 symmetrically-distinct orderings on the FCC lattice selected by an iterative approach. Each configuration was initialized ferromagnetically during the DFT calculations. We started with all known FCC based ground states as well as all symmetrically distinct orderings of Co and Pt over FCC within supercells containing up to 6 atoms. A cluster expansion was fit to these energies and subsequently used to search for low energy configurations in larger supercells. The energies of configurations in the larger supercells that were predicted to be below or close to the convex hull with the cluster expansion were then calculated with DFT and included in a new fit. The cluster expansion was iteratively improved until no new ground states were predicted. The resulting set of structures included: (i) all unique supercells up to 6 atoms, (ii) all supercells up to 8 atoms with platinum composition between 25\% and 75\%, and (iii) long-period superstructure orderings involving (001) layers of Co and Pt stacked in various tilings in supercells with up to 15 atoms (1 $\times$ 1 $\times$ \textit{n} primitive cells). Additional magnetic configurations were tested; anti-ferromagnetic (AFM) and mixed/ferrimagnetic orderings were explored (details of the orderings are given in the Supplemental Material, Fig. S1\cite{supp}), and non-collinear magnetism with spin-orbit coupling (SOC) corrections was separately included. The formation energies of all these configurations, calculated using HCP Co and FCC Pt as reference states, are shown in Fig.~\ref{fig1}.

The set of ground states predicted by PBE (red circles/lines in Fig.~\ref{hull}, corresponding structures in Fig.~\ref{struct}) contain a variety of ordered phases at differing compositions. Most prominent is the $\beta_2$ CoPt$_2$ ordering, characterized by alternating single (001) layers of Co and pairs of (001) layers of Pt. The L1$_0$ structure, observed experimentally through a wide range of compositions\cite{Inden1983, Leroux1988}, is predicted to be stable only in a very narrow chemical potential range as a result of the small difference between the slopes of the tie lines connecting L1$_0$ CoPt to Co$_2$Pt and Co$_2$Pt$_3$. The experimentally observed L1$_2$ CoPt$_3$ ordering is entirely absent from the set of ground states, excluded by the depth of the common tangent between the CoPt$_2$ and FCC Pt phases. A large number of long-period superstructures, characterized by different arrangements of (001) layers of Co and Pt and with a preference for Pt-Pt layer pairs (similar to $\beta_2$) to account for deviations in stoichiometry (instead of anti-site defects on a L1$_0$ supercell), are present along or within 5 meV of the tangent between CoPt and CoPt$_2$.

These enthalpies and ground states match the results of Chepulskii \textit{et al}.\cite{Chepulskii2011} and Hart \textit{et al}.\cite{Hart2014}, although an order of magnitude more configurations have been considered here. While both prior works have reported a stable D0$_{19}$ (HCP) phase at Co-rich compositions\cite{Hart2014, Chepulskii2011}, we restricted our focus to FCC-like superstructures because we are only interested in equiatomic and majority-Pt alloys (i.e., $0.25 < x_{Pt} < 0.75$), which all adopt FCC-based orderings experimentally\cite{Kim2011}.

The calculated formation energies of L1$_0$ CoPt and L1$_2$ CoPt$_3$ are approximately 50 meV above (less stable) the measured formation enthalpies at 914 K\cite{Oriani1962}, i.e., an error of nearly 40\%. The lattice parameters, however, are within $<$1\% of experimental measurements\cite{Leroux1988}. Alternative magnetic configurations and spin-orbit contributions did not result in any further lowering of the formation energies of the ground states and \textit{increased} the difference between calculated and measured formation enthalpies [blue triangles, green squares, and purple pentagons in Fig.~\ref{hull}, respectively].

\begin{figure}[h]
\subfigure{\label{ECIs} \includegraphics[width=.45 \textwidth]{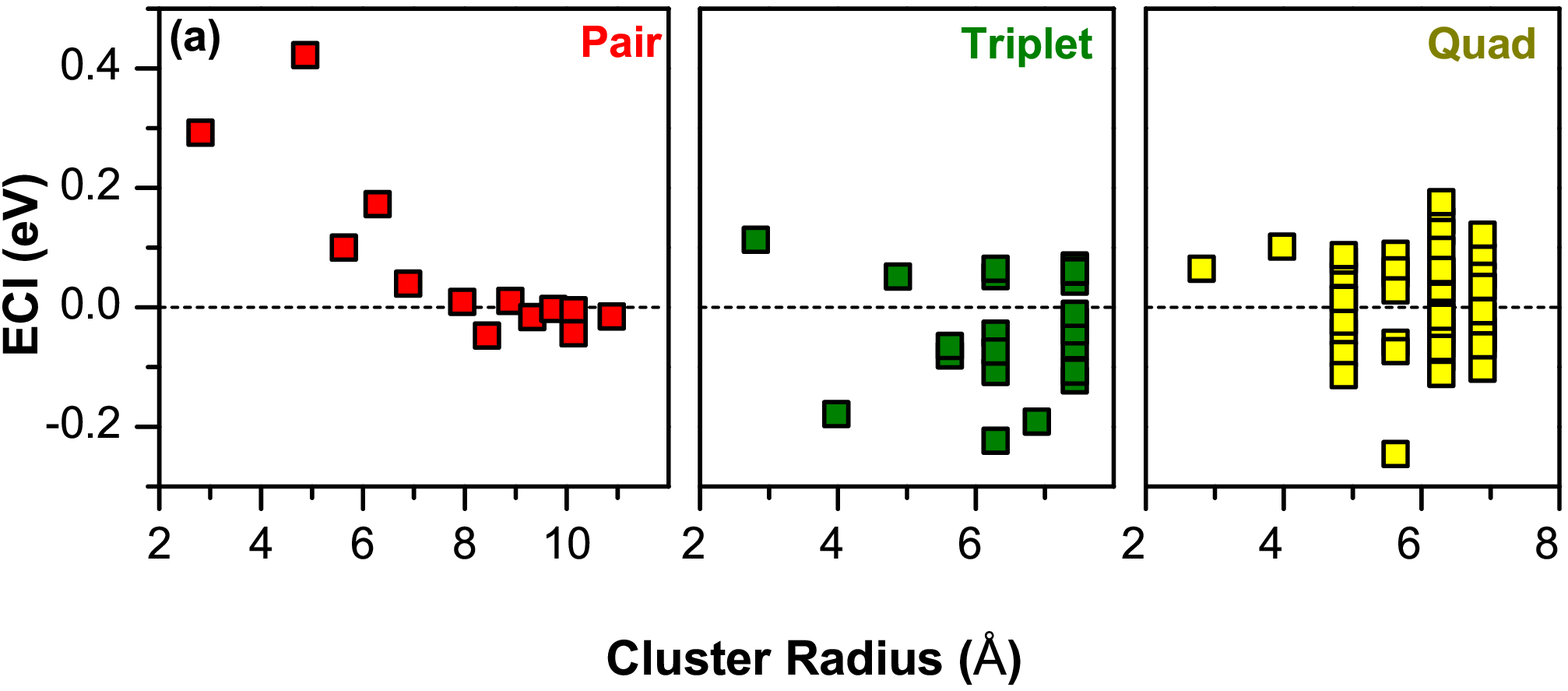}}
\subfigure{\label{phases} \includegraphics[width=0.45 \textwidth]{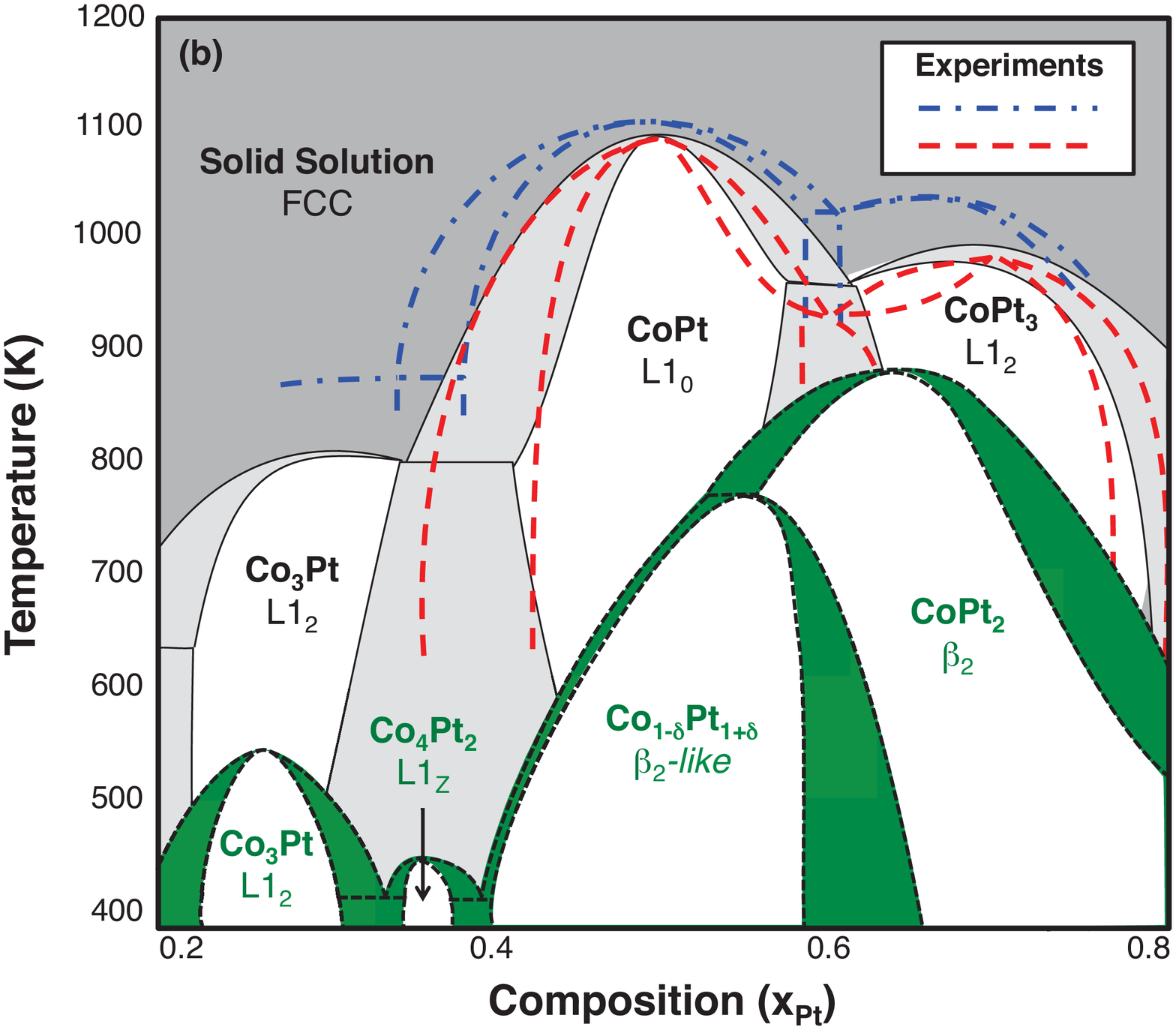}}
\caption{(color online) (a) Effective cluster interactions of the PBE cluster-expanded Hamiltonian, grouped by cluster type. The empty and point clusters have ECIs of 20.5 meV and 473 meV, respectively. (b) The temperature-composition phase diagrams for Co-Pt from experiments (dashed red, blue lines\cite{Inden1983,Leroux1988, Sanchez1989}), our results with PBE (green), and CALPHAD (solid black lines\cite{Kim2011}). Green shading indicates two-phase regions.\label{ECI_phases}}
\end{figure}

The Co-Pt system forms a FCC-based solid solution at high temperatures and a variety of FCC-based ordered structures upon cooling. The contributions of configurational entropy play an important role in determining phase stability when increasing temperature. We fit\cite{Puchala2013} the coefficients of a cluster expansion to the 1469 formation energies using a genetic algorithm\cite{Hart2005} to determine the optimal basis set. The resulting fit has 89 effective cluster interaction (ECIs) coefficients corresponding to pair, triplet, and quadruplet clusters [Fig.~\ref{ECIs}]. The root-mean-squared error of the fit was 3.3 meV, and the cross validation score (using leave-one-out cross validation) was 3.6 meV.

The PBE-based phase diagram [Fig.~\ref{phases}], determined using semi-grand canonical Monte Carlo, shows a very wide stability region for $\beta_2$ CoPt$_2$. Although this ordering also has the highest order-disorder temperature, the transformation is predicted to occur hundreds of degrees lower than the experimental transition temperatures of L1$_0$ CoPt and L1$_2$ CoPt$_3$\cite{Inden1983,Leroux1988, Sanchez1989}. The L1$_2$ CoPt$_3$ structure does not appear at all in the calculated phase diagram, and the region surrounding $x_{Pt} = 0.5$ consists of a continuum of defected incommensurate long-period superstructures up to the peritectoid temperature, decomposing into a mixture of solid solution and $\beta_2$ CoPt$_2$. We found no evidence for the stabilization of defected L1$_0$ CoPt (i.e., single planes of Co and Pt with a random distribution of anti-site defects); even at compositions close to $x_{Pt} = 0.5$, the structure instead resembled $\beta_2$-like Co$_3$Pt$_5$ with anti-site defects concentrated in the paired Pt-Pt (001) layers. The phase diagram was not explored below 400 K, nor were compositions below $x_{Pt} = 0.2$ or above $x_{Pt} = 0.8$ explored because these regions were not emphasized when fitting the cluster expansion.

Overall, the resulting phase diagram is inconsistent with the high temperature observations of L1$_0$ CoPt and L1$_2$ CoPt$_3$. These phases have been well characterized in the literature and have important differences in diffraction patterns from the long-period superstructures predicted by PBE\@. Experimental characterization of the L1$_0$ phase has historically relied upon the [001] superstructure peak\cite{Newkirk1951,Rudman1957,Leroux1988}, a peak that is absent in the predicted long-period superstructure orderings.  L1$_{2}$ CoPt$_{3}$, having cubic symmetry, also has very different diffraction patterns from tetragonal $\beta_2$ CoPt$_{2}$ or any of its derivatives.

% \subsubsection{Phonons from PBE}

While the calculated phase diagram only accounts for configurational excitations at finite temperature, vibrational and magnetic excitations may also play a role in determining the relative stability of different orderings. To explore vibrational free energies, we performed phonon calculations for FCC Co, FCC Pt, L1$_{0}$ CoPt, $\beta_{2}$ CoPt$_{2}$ and L1$_{2}$ CoPt$_{3}$ within the quasi-harmonic approximation. However, even at 914 K, L1$_{2}$ CoPt$_{3}$ never emerges as a stable phase relative to $\beta_{2}$ CoPt$_{2}$ and FCC Pt. Furthermore, the stability of L1$_{0}$ CoPt relative to $\beta_2$ CoPt$_2$ and L1$_2$ Co$_3$Pt does not increase markedly with increasing temperature. This suggests that rigorous inclusion of vibrational degrees of freedom, together with configurational degrees of freedom using coarse-graining schemes\cite{Asta1993, Ceder1993}, are unlikely to qualitatively alter the calculated phase diagram of Fig.~\ref{phases}.

Thermal excitations of magnetic moments have previously been shown to be important in affecting phase stability in Co-Pt alloys\cite{Inden1983, Sanchez1989}. In pure Co, entropic contributions arising from spin-spiral excitations have a strong influence on the HCP/FCC transformation temperature\cite{Uhl1996}. At Pt-rich compositions, the ferromagnetic to paramagnetic Curie temperatures occur below the order-disorder transition temperatures of both L1$_0$ CoPt and L1$_2$ CoPt$_3$~\cite{Inden1983, Leroux1988, Kim2011}. Hence, magnetic disorder, neglected in the calculation of our phase diagram, will likely play a role in determining the precise order-disorder transition temperatures. An in-depth study of the effects of magnetic entropy on the phase diagram are beyond the scope of this study. However, below the Curie temperatures contributions to the free energy from magnetic entropy are expected to be small. While we do not know the Curie temperature of the $\beta_2$ phase, those of L1$_0$ and L1$_2$ are above all of our calculated solid solution transformation temperatures. In the case of the $\beta_2$ phase, there are two possibilities: (i) CoPt$_2$ remains ferromagnetically ordered up to the solid solution transformation temperature, or (ii) CoPt$_2$ becomes paramagnetic below the solid solution transformation temperature. In scenario (i) our phase diagram should be negligibly impacted by the inclusion of magnetic entropy, while in scenario (ii), magnetic entropy will only further \textit{stabilize} the $\beta_2$ phase with respect to the $L1_0$ and $L1_2$ phases. We therefore expect that inclusion of magnetic excitations will not rectify the disagreement between finite temperature predictions and experimental observations. 

The over-stabilization of $\beta_2$ and other long-period superstructures with paired (001) Pt-Pt planes can be attributed to over-delocalized electron charge densities in PBE\@. Generalized gradient approximations (GGAs) can perform poorly in transition metals\cite{Furche2006,Ruban2008,Cramer2009} where the significance of the localized \textit{d}-orbitals comes into conflict with the orbital-less approach of GGA\@. Any functional treatment of the electron density leads to an electron interacting with its own potential, the self-interaction error, which causes excessive delocalization of the total charge. Additionally, the GGA functional can not energetically differentiate between occupied and unoccupied bands, leading to incorrect predictions for orbital/band occupation and splitting\cite{Mori-Sanchez2008,Cohen2008}. Because both the energy levels and occupations of the \textit{d} orbitals are incorrect, the Co 3\textit{d} and Pt 5\textit{d} bands cannot hybridize, losing significant enhancement of the magnetic moment\cite{Sakuma1994,Uba2001,Sipr2008}. Stabilization of a magnetic ground state, however, is a driving force in choosing the thermodynamic ground states in Co-Pt\cite{Karoui2013}, and ferromagnetic effects drive the asymmetry in the phase diagram with respect to the Co-rich and Pt-rich L1$_2$ phases\cite{Inden1983, Sanchez1989}.

\subsection{Comparison with Hybrid Functional HSE06}

Using a screened form of Hartree-Fock exchange, Zhang \textit{et al.}\cite{Zhang2014} were able to recover the experimental ground-states of Au-Cu. The parametrization of the hybrid functional developed by Heyd \textit{et al.} (HSE06)\cite{Krukau2006,Heyd2003} reduces the self-interaction error akin to Hartree-Fock theory\cite{Henderson2011}, while avoiding the associated singularity in occupation at the Fermi level. Since HSE06 explicitly includes orbitals, \textit{d}-orbital hybridization can be recovered, e.g., as in Au-Cu. The functional is, however, limited by its computational expense\cite{Sorouri2006} and accuracy\cite{Cramer2009, Janthon2014}.

To examine the performance of the HSE06 functional, all of the PBE ground states, as well as L1$_2$ CoPt$_3$, FCC Co and Pt, and HCP Co were recalculated [pink hexagons/lines in Fig.~\ref{hull}]. The results show qualitative improvement: L1$_2$ CoPt$_3$ is predicted as a ground state, and L1$_0$ CoPt is substantially more stable relative to  L1$_2$ Co$_3$Pt and $\beta_2$ CoPt$_2$. However, the enthalpy and lattice parameter errors increase by an order of magnitude (200--300 meV and 2.5--4.5\%, respectively) and $\beta_2$ remains as a ground state. These results are not surprising: HSE06 is known to severely overestimate exchange splitting in itinerant magnetic systems\cite{Paier2006a, Stroppa2007, Janthon2014}. The default screening parameter in the HSE06 formalism ($\omega = 0.2$ \AA$^{-1}$, $r_{screen} = 10$ \AA) results in an effective screening length an order of magnitude greater than the length scales ($r_{screen} = 0.24-0.26$ \AA) for screening in bulk Pt or Co metals\cite{Rose1983}, introducing spurious interactions between orbitals at different sites. This is further verified by stabilization of a \textit{ferromagnetic} ground state for Pt in both this work and Ref.~\onlinecite{Tran2012} over the experimentally observed nonmagnetic state, and an overestimation of magnetic moments with HSE06 for \textit{all} experimentally studied structures.

% \subsubsection{Density of States}

\begin{figure}[h]
\includegraphics[width=0.45 \textwidth]{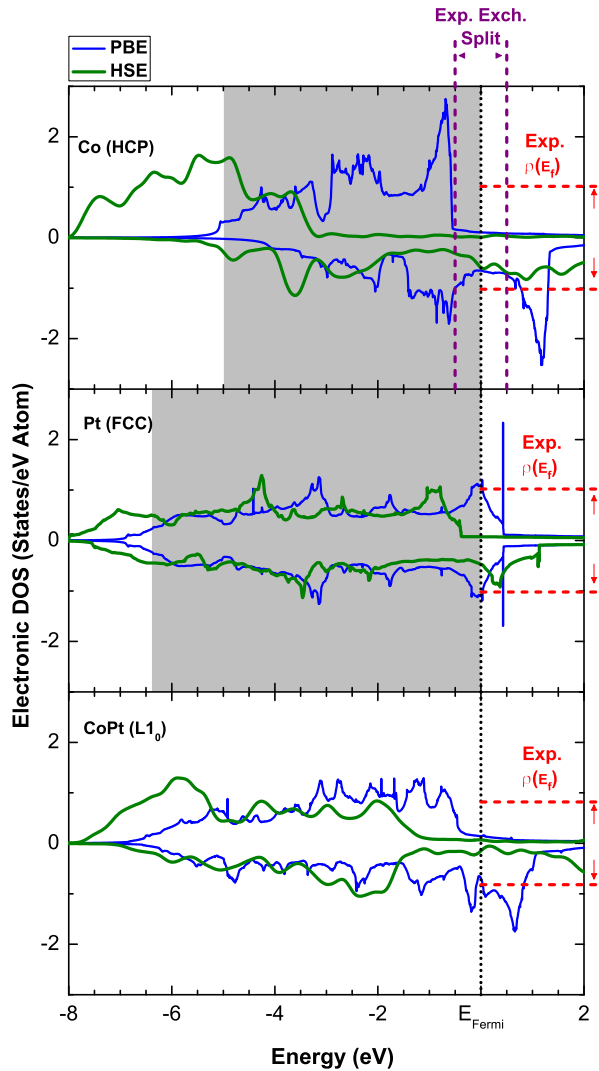}
\caption{(color online) \textit{d}-orbital density of states for HCP Co (top), FCC Pt (middle) and L1$_0$ CoPt (bottom) calculated from DFT, using both the PBE (blue line) and HSE06 (green line) functionals. Gray regions show occupied \textit{d}-band width from photoemission results\cite{Heimann1977, Lin1971}, red dotted lines show occupation at the Fermi level predicted from the heat capacity\cite{Cheng1960, Shoemake1968, Kuntzler1981}, and purple dotted lines show exchange splitting predicted from experiment [$\Delta E$ between $N{(E)}_{\max}$ for majority and minority spins] for Co\cite{Eastman1980}.\label{DOS}}
\end{figure}

The shortcomings of both the PBE and HSE06 functionals are more easily visualized by comparing the calculated density of states (DOS) for Co-Pt (Fig.~\ref{DOS}) with data from experiments. The width of \textit{d}-band states, as measured by photoemission\cite{Heimann1977, Lin1971}, are shown in grey in Fig.~\ref{DOS}, while the electron occupation at the Fermi level can be inferred from low temperature heat capacity experiments\cite{Cheng1960, Shoemake1968, Kuntzler1981}, shown as the height between the red dashed lines. In the case of HCP cobalt, PBE predicts a \textit{d}-band width in good agreement with experiment, but the occupation at the Fermi level is underestimated. The excessive delocalization provided by self-interaction errors in PBE smears the electron density towards a more even distribution between orbitals, and this effect is magnified by the inability of PBE to (energetically) distinguish between occupied and empty orbitals. This latter point is exemplified by a difference between the calculated exchange splitting (energy difference between maxima in spin-up and spin-down DOS) and the experimentally determined width, indicated in purple. Unsurprisingly, PBE performs well in FCC platinum, where the \textit{d}-orbitals are fully occupied and differentiation between sub-shells does not matter. Though we were not able to find suitable experimental analysis of the L1$_0$ CoPt \textit{d}-bands, Fig.~\ref{DOS} shows an unexpectedly wide \textit{d}-band and low electronic occupation at the Fermi level in CoPt, similar to Co.

% \subsubsection{Discussion}

Contrary to expectation, HSE06 worsens the degree of electronic delocalization in Co even as the self-interaction error should be reduced; this can be attributed to the unrealistic (for Co) screening length standardized in the HSE06 functional. In Pt, a very large fictitious exchange splitting is introduced, explaining the large magnetic moment commented upon previously. Here, an electron has been promoted from the 5\textit{d} orbitals in PBE to the 6\textit{p} orbitals in HSE06 to increase the net number of unpaired spins. The over-stabilization of a magnetic ground state also reduces the energy of the \textit{d}-orbitals in all materials; we can see this in Fig.~\ref{DOS} where the occupation at the Fermi level drops to between one half and one quarter of the PBE values. These results both highlight the failures of the PBE and HSE06 functionals and offer insight as to \textit{why} these methods describe the Co-Pt binary so poorly.

\subsection{Additional Functionals}

% LSDA+U incorporates an additional on-site Coulomb and exchange term akin to the Hubbard model, expressed as "coupling parameters" $U$ and $J$ with either both\cite{Liechtenstein1995} or the difference ($U - J$)\cite{Dudarev1998} entering into the final calculation. The coupling parameters are site-specific and are usually determined semi-empirically to agree with existing experimental data\cite{Himmetoglu2014}, limiting the use of this approach as a true \textit{ab-initio} method. In an attempt to fit both the experimental family of ground states and associated $c/a$ lattice parameter ratios, 

Additional functionals were explored, though not in-depth, once initial results indicated a qualitatively similar set and arrangement of ground states to the results using PBE\@. The Tao-Perdew-Staroverov-Scuseria\cite{Tao2003} meta-GGA functional and its revised variant\cite{Perdew2009}, as well as the Minnesota meta-GGA functional M06-L\cite{Zhao2006}, were tested, but retained $\beta_2$-like structures and excluded L1$_2$ CoPt$_3$ from the set of ground states. PBESol\cite{Csonka2009} was investigated but also produced similar results to PBE\@. HSESol\cite{Schimka2011}, and various parametrizations of the HSE functional (in $\alpha$ and $\omega$) were tested only on L1$_0$ CoPt, L1$_2$ Co$_3$ Pt and CoPt$_3$, and $\beta_2$ CoPt$_2$, but yielded decreasingly tiny occupations at the Fermi level as well as worsening energetics and a deepening of the $\beta_2$ enthalpy with respect to other ground states.

The importance of localized \textit{d} electrons in Co-Pt, which has been incorrectly treated in both PBE and HSE06, motivated the examination of the LSDA+U method\cite{Anisimov1991}. We explored a two-dimensional grid of $U$'s of 1.0--4.0 eV in Co, and 0.0--4.0 eV in Pt, in 0.1 eV steps using the rotationally-invariant method of Dudarev \textit{et al.}\cite{Dudarev1998}. Unfortunately (though perhaps not unsurprisingly), the results showed no set of $U$'s that simultaneously produced the correct set of ground states, matched the experimental enthalpies (within a range of $\pm 50\%$), and matched the experimental $c/a$ ratio (within a range of $\pm 50\%$). The linear-response approach to determine coupling parameters of Cococcioni and Gironcoli\cite{Cococcioni2005} was also used, but the resulting $U$'s produced similar unsatisfactory results.

\section{Conclusion}

In summary, we have shown that the anomalous ground states predicted in Co-Pt using the PBE functional result in equally anomalous phase behavior at elevated temperature, wholly inconsistent with the body of experimental literature. By examining our results in the greater context of the \textit{known} shortcomings of PBE, we can characterize the modes of failure, attributing the stabilization of the $\beta_2$ CoPt$_2$ ground state to self-interaction and occupation errors inherent in the functional. Although DFT performs exceptionally well in a wide variety of inorganic systems, caution must be used when predictions appear inconsistent with experiment. By using rigorous statistical mechanical approaches, experimental results can be meaningfully compared with zero kelvin predictions in both qualitative and quantitative fashions. Unfortunately, when one method falls short, it is not always sufficient to move up the ``Jacob's ladder'' of XC functionals\cite{Perdew2001}: the HSE06 functional merely trades one set of inaccuracies for another. Based on both our analysis and the existing literature, we believe that sibling systems (e.g., Fe-Pt, Ni-Pt, Fe-Ni) will yield similar results. Though the prospect of a one-size-fits-all DFT-based approach to predicting phase diagrams is appealing, our analysis of the Co-Pt system highlights the need to review zero kelvin electronic results in a finite-temperature and thermodynamically meaningful fashion.

\begin{acknowledgments}
We thank Min-Hua Chen and Dr.\ John C. Thomas for use of their phonon calculation code. We also thank Dr.\ Paul Weakliem for help with computational facilities. Finally, we thank Min-Hua Chen, Anirudh Natarajan, and John Goiri for many helpful discussions and assistance with performing calculations. This work and Elizabeth Decolvenaere were supported by the MRSEC Program of the National Science Foundation under Award No. DMR-1121053. Simulations were performed using resources from the Center for Scientific Computing in the CNSI and MRL, funded by National Science Foundation MRSEC (DMR-1121053), National Science Foundation CNS-0960316, and Hewlett Packard.
\end{acknowledgments}

\bibliography{CoPt.bib}

\end{document}